# Association of stroke lesion distributions with atrial fibrillation detected after stroke

**Yiming Chen, MD#[1], Sihui Wang, MD#[2], Dong Xu, MD, PhD* [3]**

1 Department of cardiac surgery, Beijing Tiantan Hospital, Capital Medical University, Beijing, China

2 Department of Radiology, Beijing Tiantan Hospital, Capital Medical University, Beijing, China

3 Department of cardiac surgery, Beijing Tiantan Hospital, Beijing, China

# YimingChen and Sihui Wang contributed equally to this work.

*Correspondence author: Dong Xu, MD, Ph.D.

   Correspondence: Department of cardiac surgery, Beijing Tiantan Hospital, No.119, The West Southern 4th Ring Road, Fengtai District, Beijing, 100070, China.

E-mail address: Dr_DXu@163.com

**Abstract**

*Background*

Atrial fibrillation is often missed by traditional intermittent electrocardiogram monitoring after ischemic stroke due to its paroxysmal and asymptomatic nature. The knowledge of the unique characteristics of the population with atrial fibrillation detected after stroke (AFDAS) enables more ischemic stroke patients to benefit from more aggressive anticoagulation therapy and AF management.

*Method*

This is an observational, retrospective, MRI imaging-based single-center study. Patients with AFDAS were matched in 1:3 ratio with patients without AF (NoAF)and patients with known AF before stroke(KAF) in PSM model based on age, gender, and time from stroke onset to admission. Multivariate logistic models were used to test the association of MRI-based stroke lesion distribution, other clinical parameters and AF. A backward stepwise elimination regression was conducted to identify the most important variables.

*Results*

Compared to the NoAF group(n=103), the patients with AFDAS (n=42) had more cortical involvement(p=0.016), as well as temporal(p<0.001) and insular lobes(p=0.018) infraction. After performing a backward stepwise elimination model in regression analysis, the temporal lobe infraction(OR 0.274, 95%CI 0.090-0.838, p=0.023) remained independently associated with the detection of AF. Compared to the KAF group(n=89), LAD(OR 1.113, 95%CI 1.022-1.211, p=0.014), Number of lobes infarction(p=0.012), 3-lobes involvement(OR 0.177, 95%CI 0.056-0.559, p=0.003), and left hemisphere lobe involvement(OR 5.966, 95%CI 2.273-15.817, p<0.001)) were independently associated with AFDAS and KAF.

*Conclusions*

Ischemic stroke patients with AF detected after stroke present more temporal lobe infraction and cortical involvement. These lesion distribution characteristics with clinical characteristics together may help in stratifying patients with long-term cardiac monitoring after stroke.

## 1. Introduction

Atrial fibrillation (AF) is the most common tachyarrhythmia in the world[1], with approximately 20-30% of stroke and TIA patients having known atrial fibrillation (KAF) [2,3]. Due to the occult and paroxysmal nature of AF, accounting for 24% of all stroke and TIA patients with AF remain undetected on admission ECG[4]. The stroke risk in AF patients is 3-5 times higher compared with the general population[5], and the strokes related to AF have been described as more severe than strokes due to other sources[6], while this risk can be obviously reduced by oral anticoagulants[7]. Suspected cardiac ischemic stroke patients who are not found to have AF are usually advised to use antiplatelet agents for stroke secondary prevention, which have been shown to have higher stroke recurrence rates than patients using NOAC[7,8]. Typically, Several strategies including long-term Holter[9], internal cardiac monitor(ICM)[10], and other external monitoring devices[11,12] are used to increase the detection rate of paroxysmal AF in suspected stroke patients. Even though the whole cardiac monitoring duration is fewer than 30 days also has been shown to have a higher AF detection rate than conventional ECG.

As a novelty clinical concept, AFDAS is considered to be newly diagnosed atrial fibrillation detected through the aforementioned long-term monitoring methods[13,14]. Compared to patients with known AF before the stroke, the AFDAS population has different demographic characteristics, as well as risks and outcomes including stroke recurrence [15]. However, most of these data and studies come from Western populations and may not represent the reality in other regions. Furthermore, the causality between AFDAS and TIA or stroke still remains unclear, as stroke per se could cause heart disease including arrhythmia and myocardial injury, which is called stroke-heart syndrome[16,17]. The development of left atrial endothelial dysfunction, inflammation, and fibrosis early after the induction of insular stroke has been demonstrated in a rat model[18]. Recent research has found a definite relationship between the stroke lesion shape and AFDAS[19], and the relationship between the involvement of the insular lobe, cortical and AFDAS had also been reported[19-22]. Additionally, previous studies found various bio-markers showed potential association with AFDAS, including brain natriuretic peptide(BNP), and high-density lipoprotein cholesterol(HDL) [23].

Therefore, unique demographic characteristics, bio-markers, and MRI imaging lesion patterns may be valuable for stratifying suspected cardiogenic ischemic stroke patients. Here, based on real-world experience among the Chinese population, we describe the unique demographic characteristics and MR imaging-based stroke lesion distribution patterns of AFDAS, as well as potential predictors of AFDAS, by comparing with the KAF patients and those stroke patients without AF.

## 2. Materials and Methods

### 2.1. Study Design

In this observational cohort study, we retrospectively analyzed the ischemic stroke patients who underwent long-term electrocardiogram (ECG) monitoring at our centre from September 2020 to September 2022. By comparing the Atrial Fibrillation Detected After Stroke(AFDAS) group, which was diagnosed newly AF through long-term ECG monitoring after admission, with the NoAF group, which had no AF detected in the same population, as well as the reference group consisting of patients with known AF (KAF) prior to stroke, we explored the unique demographic characteristics and other predictors of the AFDAS population.

Considering age and gender as risk factors for atrial fibrillation and common elderly diseases including hypertension and diabetes, as well as the potential impact of stroke onset time on patients' physiological parameters. After excluding patients whose first stroke occurred more than 14 days prior to admission, we matched the AFDAS, KAF, and NoAF groups in a 1:3 ratio based on age, sex, and occurrence time to admission.

Both the AFDAS and NoAF groups in this study were from the same population who underwent long-term cardiac monitoring using wearable patch cardiac monitoring devices. The KAF group was considered as a reference group to give more weight to our findings. This group consisted of patients who were diagnosed with ischemic stroke from unknown sources as their primary diagnosis at our centre from February 2019 to February 2023 and had a definite history of AF prior to stroke onset.

*2.2. External wearable patch monitoring devices*

The wearable patch cardiac monitor system (Ambulatory Electrocardiography System PE-C, Prudence Medical Technologies, Shanghai, China) consists of a recorder, software, and record analysis server and was approved commercially available by China Food and Drug Administration (CFDA). This device had a memory capable of recording ECG for up to 30 days using a simulative one-channel with a 256 Hz sampling rate. Technical parameters include dynamic range ($\pm 5$ mV), input impedance ($>10$ M$\Omega$), common mode rejection ($\geq 60$ dB), gain accuracy ($\leq 10\%$), gain stability ($\leq 3\%$ over a 24-hr period), system noise ($<50$uVp-v), and frequency response (0.67–55 Hz) were set as the factory setting. The recorder was applied over the left pectoral region of the patient's chest, and the collected data were continuously transmitted to the server for analysis. The final diagnosis was finally admitted by capable physicians in our institute through cardiac monitor data.

*2.3. Imaging data collection*

All patients were imaged using a 3T Siemens Verio scanner (Siemens Medical Systems, Erlangen, Germany) equipped with a 12-channel head coil. Scanning sequences included transversal DWI with diffusion-sensitizing gradients in at least three orthogonal directions and b values of 0 and 1000 mm2/s together with transversal gradient recalled echo, T1/T2-weighted sequences（T1WI/T2WI）, and fluid-attenuated inversion recovery (FLAIR) sequences. Part of the scanning parameters were as follows: T1WI: TR(repetition time): 1900 ms; TE(echo time): 8.6ms; slice thickness, 5 mm; slices, 24; field of view, 512mm×496 mm; T2WI：TR, 4500 ms; TE, 99ms; slice thickness, 5 mm; slices, 24; field of view, 640mm×640 mm. Patients who underwent solely cranial CT or CT angiography were excluded, as were patients with MRI scans confounded by motion or other artefacts. Imaging analysis was conducted using an in-house PACS system. The lesion was delineated independently by two neuroradiologists based on DWI and T1WI sequences using ITK-SNAP 4.0, and the two participants reached a consensus on disputed parts after consultation.

*2.4 Lesion Overlay map*

To clearly and directly visualize the distribution of lesions, we collected the postoperative MRI scans for all the patients who had lobes involvement. The resected brain area was segmented on T1- or T2-weighted images by an experienced radiologist to generate a lesion mask via the Medical Imagining Interaction Toolkit (MITK) (www.mitk.org). Then the lesion masks and corresponding anatomical images were normalized into the standard Montreal Neurological Institute (MNI) space via SPM12. Next, the radiologist performed a reconfirm procedure for more accurate matching to the original lesion range after normalization. Finally, the normalized and rechecked lesion masks were generated for lesion overlap using MRIcroGL (https://www.mccauslandcenter.sc. edu/mricrogl).

## 2.4. Statistical analysis

Normal distribution, abnormal distribution continuous, and categorical variables were expressed as mean±SD, median(IQR), and counts (percentage) respectively. Comparisons of baseline characteristics, bio-markers level and lesion distribution across the cohort were compared using the Kruskal-Wallis test, Wilcoxon rank-sum test, Chi-square tests or Fisher exact test where appropriate. Then the logistic regression models were fitted was fitted including all intergroup significantly different paraments found by the aforementioned methods. This regression model was further refined by the backward stepwise elimination algorithm. During every step, variables that did not reach the significance level of 0.1 were eliminated from the model starting with the least significant variable. 2- tailed $P <0.05$ was considered statistically significant. SPSS Version 26 (SPSS Inc. Chicago, IL, USA; released 2013) was used for statistical analysis.

**Figure 1. The chart flow for the enrollment of patients**

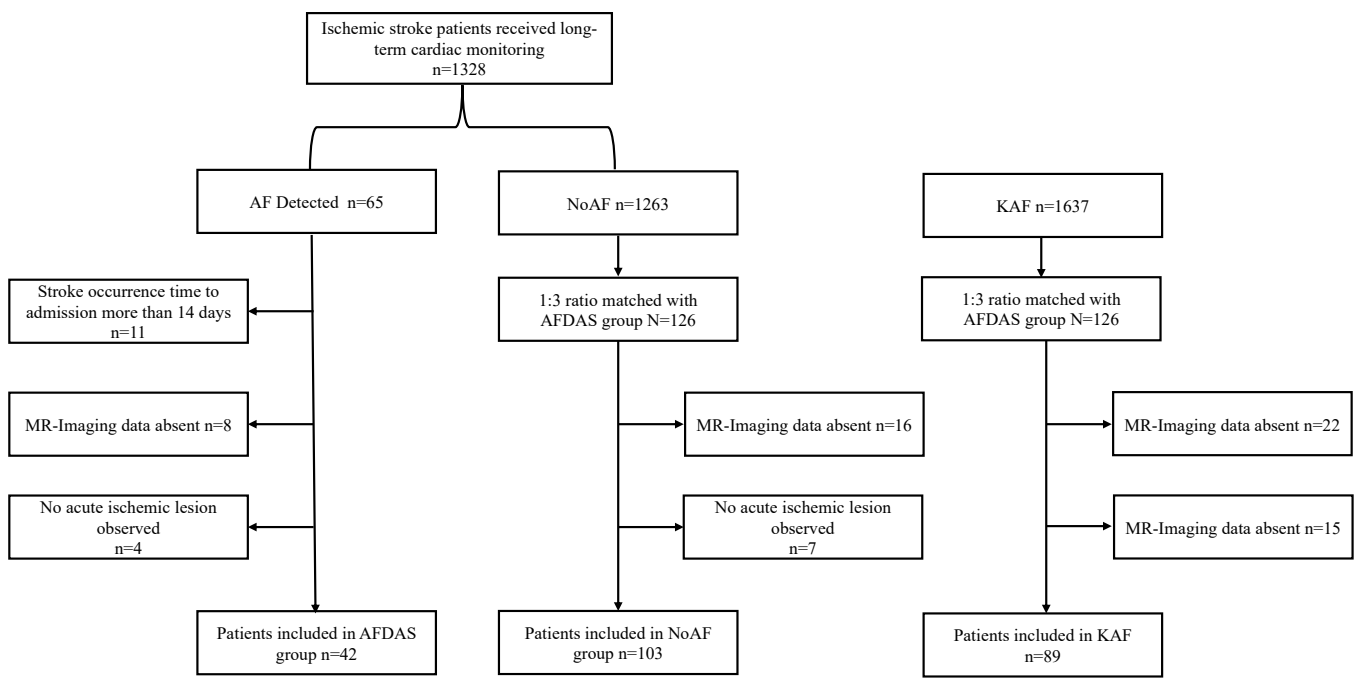

## 3. Results

Between September 2020 and September 2022, our centre performed long-term cardiac monitoring on a total of 1328 suspected ischemic stroke or transient ischemic attack patients, with an average duration of $7.44 \pm 3.61$ days. Of these, 65 patients were diagnosed with atrial fibrillation after an average of $3.54 \pm 1.32$ days of monitoring. 11 patients were excluded as the onset of stroke to admission more than 14 days, 8 and 4 patients were excluded due to lack of MRI imaging data and failure to detect acute ischemic stroke lesion under the MRI, respectively. Ultimately, the AFDAS group included 42 patients. Propensity score matching (PSM) was performed for age, gender, and the onset of stroke to admission in a 1:3 ratio between the AFDAS group with 1263 patients who were confirmed not to have AF by long-term cardiac monitoring, as well as 1637 patients admitted with ischemic stroke as the primary diagnosis and a definite history of AF from March 2019 to March 2023. Ultimately, 103 patients were included in the NoAF group, and 89 patients were included in the KAF group. The patient chart flow can be seen in Figure 1.

Table 1 shows the baseline demographic results of the three patient groups. The mean age of the AFDAS group is 68.64 ± 10.11 years and only 19.05%(n=8) of patients were female, as well as the onset of stroke to admission is 2 (3) days. The three aforementioned factors had no significant differences compared to the NoAF and KAF groups. The proportion of patients diagnosed with heart failure (HF) upon admission was significantly higher in the AFDAS group than in the NoAF group, but similar to that in the KAF group. Compared to the NoAF group, AFDAS patients exhibited lower rates of comorbidities, including hypertension, diabetes mellitus, and peripheral vascular disease. The proportion of prior stroke and the mean CHA2DS2-VASc score in the AFDAS group were the lowest among the three groups, while the former showed significant statistical differences between the AFDAS group and the other two groups, the CHA2DS2-VASc score only showed statistical differences between the AFDAS and KAF groups.

**Table 1.** Patient baseline characteristics

|  | NOAF | p value | AFDAS | p value | KAF |
|---|---|---|---|---|---|
| Femal | 18(17.48) | ns | 8(19.05) | ns | 23(25.84) |
| Age | 68.44±9.67 | ns | 68.64±10.11 | ns | 69.974±10.29 |
| Stroke onset time | 2(3) | ns | 2(3) | ns | 2(2) |
| BMI | 25.064±4.28 | ns | 25.714±4.12 | ns | 24.934±3.984 |
| HF | 15(14.56) | 0.049 | 12(28.57) | ns | 28(31.46) |
| Hypertension | 88(85.44) | 0.023 | 29(69.05) | ns | 70(68.65) |
| DM | 52(50.49) | 0.175 | 16(38.10) | ns | 33(37.08) |
| Vascular diseases | 92(89.32) | 0.003 | 29(69.05) | 0.095 | 73(82.02) |
| Prior stroke history | 27(26.21) | 0.026 | 4(9.52) | 0.023 | 24(26.97) |
| Cardiac diseases | 13(12.62) | ns | 8(19.05) | 0.021 | 35(39.32) |
| Smoke | 58(56.32) | ns | 24(57.14) | ns | 44(49.44) |
| Alcohol | 41(39.81) | 0.166 | 22(52.38) | ns | 38(42.70) |
| CHA2DS2-Vasc score | 5.324±1.12 | ns | 5.074±1.39 | 0.042 | 5.614±1.364 |
| <4 | 4(3.88%) | 0.025 | 6(14.29%) | ns | 8(8.99%) |
| =4-6 | 82(79.61%) | ns | 30(71.43%) | ns | 60(67.41%) |
| >6 | 17(6.80%) | ns | 6(14.29%) | ns | 21(23.59) |

ns: no significance, means p>0.20; BMI: Body Mass Index; HF: Heart failure;

DM: Diabetes mellitus;

In the comparisons of Cr, CRP, Hs-CRP, and BNP levels, the AFDAS group had significantly higher average levels than the NoAF group, and was in the intermediate level among the three groups, but there was no statistically significant difference in these variables between the KAF group and the AFDAS group (Table 2). Compared to the NoAF group, AFDAS patients had significantly higher LAD and LVD. There was an obvious increase in the LAD and LVD in the KAF group compared to the AFDAS group, but only LAD had a statistically significant difference. The average LVEF level was similar between AFDAS and NoAF groups, but in the AFDAS group is significantly higher than the KAF group.

In terms of stroke lesion distribution, compared to the NoAF group, the AFDAS group had more cortical involvement and lobes infraction (Table 2). Although the KAF group had a similar proportion of lobes infraction compared to the AFDAS group, the majority of KAF patients were diagnosed with involvement of the left lobes, while only 38.10% of AFDAS patients had left brain lobe involvement. In addition, the bilateral lobes involvement in the KAF group was also significantly higher than in the AFDAS group. Interestingly, there was no difference between the two groups in the proportion of patients with involvement of the right brain lobe. In the comparison of involvement of left, right, and bilateral brain lobes, the proportion in the NoAF group was all significantly lower than that in the AFDAS group and especially only one patient in the NoAF group had bilateral brain lobes involvement. Most of the patients in the three groups had 1-3 brain lobes infractions, among which the AFDAS group had the highest proportion of patients with 3 affected brain lobes. In terms of the distribution of affected brain lobes, the AFDAS group had more patients and a higher proportion with involvement of the frontal and insular lobes compared to the KAF group, while the AFDAS group had more patients and a higher proportion with involvement of the frontal and insular lobes compared to the NoAF group. Figure 2 shows overlap lesion maps for the AFDAS group, NoAF and KAF groups.

**Table 2**. Bio-Markers, TTE results and MRI results of ischemic lesions distribution

|  | NOAF | p value | AFDAS | p value | KAF |
|---|---|---|---|---|---|
| Cr | 66.324±15.11 | 0.060 | 70.594±13.25 | ns | 78.184±39.12 |
| LDL | 1.754±6.31 | ns | 1.124±0.24 | ns | 1.114±0.27 |
| HDL | 2.234±0.84 | ns | 2.244±1.05 | ns | 2.374±1.21 |
| CRP | 2.15(6.4) | 0.029 | 3.22(8.26) | ns | 4.26(12.31) |
| Hs-CRP | 1.91(6.07) | 0.008 | 3.09(10.25) | ns | 3.2(10.74) |
| GFR | 96.324±15.11 | ns | 97.994±13.26 | 0.169 | 90.134±23.88 |
| BNP | 65.9(81.3) | <0.001 | 173.4(335.83) | 0.118 | 231.3(357.45) |
| LAD | 33.924±3.99 | <0.001 | 37.144±3.55 | 0.002 | 39.964±6.11 |
| LVD | 44.994±4.07 | 0.030 | 46.414±5.53 | ns | 46.964±5.67 |
| LVEF | 61.364±3.65 | ns | 61.64±7.59 | 0.040 | 59.244±7.37 |
| Cortical involvement | 32(31.07%) | 0.016 | 22(52.38%) | ns | 51(57.30) |
| Lobe involvement | 45(43.69%) | 0.012 | 28(66.67%) | ns | 62(69.66) |
|   Left hemisphere | 26(25.24%) | 0.122 | 16(38.10%) | 0.001 | 62(69.66) |
|   Right hemisphere | 20(19.41%) | 0.008 | 17(39.53%) | ns | 33(37.08%) |
|   Bilateral | 1(0.97%) | 0.003 | 5(11.90%) | 0.003 | 33(37.08%) |
| Lobes involvement Number | 0(2.00) | 0.007 | 1(3.00) | ns | 1(3.00) |
|   < 3 | 91(88.34%) | 0.002 | 28(66.67%) | 0.107 | 65(73.03%) |
|   = 3 | 7(6.80%) | 0.001 | 11(26.19%) | 0.155 | 14(15.73%) |
|   > 3 | 5(4.85%) | ns | 3(7.14%) | ns | 10(11.23%) |
| Frontal lobe | 31(30.10%) | ns | 12(28.57%) | 0.074 | 40(44.94%) |
| Parietal lobe | 24(23.30%) | ns | 14(33.33%) | ns | 25(28.09%) |
| Temporal lobe | 16(15.53%) | <0.001 | 20(47.62%) | ns | 33(37.08%) |
| Insular lobe | 11(10.68%) | 0.018 | 11(26.19%) | ns | 24(26.97%) |
| Occipital lobe | 9(8.74%) | ns | 6(14.29%) | ns | 13(14.61%) |

After performing multivariate regression analysis on variables that showed significant differences between NoAF and AFDAS groups, including demographic features (HF, VD, Prior Stroke history), Cr, CRP and LAD, LVD, as well as stroke distribution (cortical involvement, lobes infraction, left lobe involvement, temporal lobe, insular lobe, 3-lobes involvement), only VD, Cr, BNP, LAD, LVD, cortical involvement were found to be significantly associated with the detection of AF (Table 3). However, after performing a stepwise elimination model, blood pressure, VD, prior stroke, CRP, Hs-CRP, BNP, LAD, and temporal lobe involvement were found to be significantly associated with AFDAS detection. Similarly (Table 4), after performing multivariate regression analysis on variables that showed differences between AFDAS and KAF groups, only left lobe involvement and 3-lobe involvement were significantly associated with AFDAS detection, while after performing a stepwise elimination model, in addition to Left and Lobe3, LAD was also significantly associated with AFDAS.

**Figure 2** Overlay map of lobe lesion density in patients without detected atrial fibrillation (AF) and with AF detected after stroke. Lesion density is displayed in the color bar on the right side.

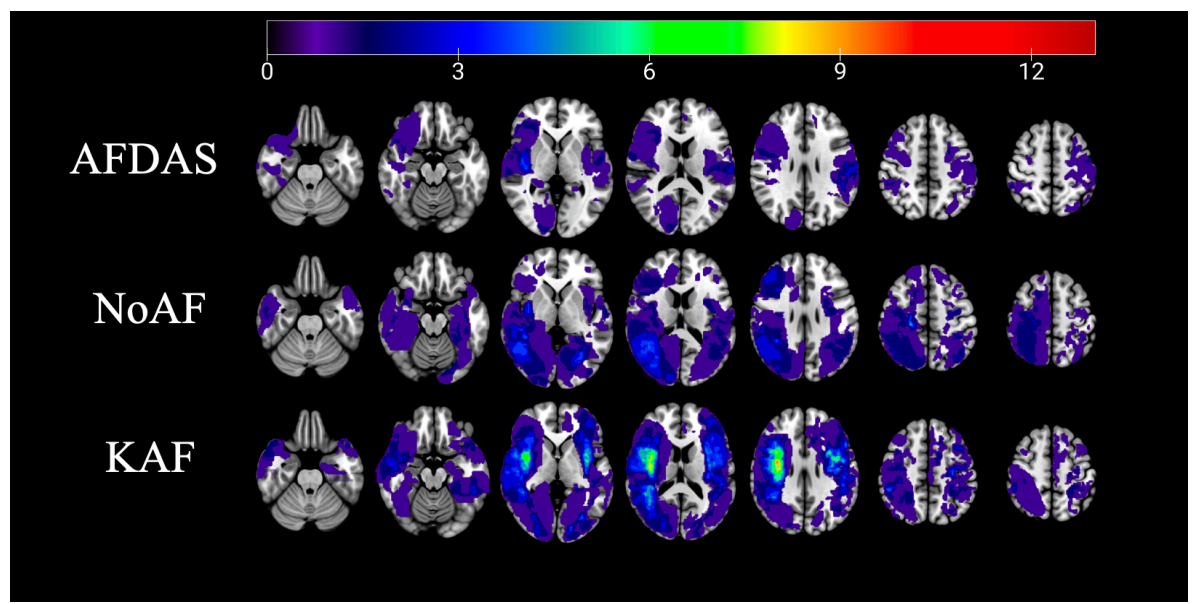

## 4. Discussion

In the chain of research on AFDAS, there are various independent risk factors in clinical population characteristics, bio-markers, ECG features, and TEE results which are reported to associate with the detection of AF post-stroke[23]. Additionally, limited research had reported more cortical and insular lobe involvement in AFDAS patients compared to those ischemic stroke patients without AF[21,22]. Furthermore, as a novelty clinical concept, AFDAS is considered as a clinical population different from KAF, with different population characteristics, risks, and outcomes[14]. Exploring the clinical features of the AFDAS population will not only enhance our understanding of the stratification of the population who could benefit from long-term cardiac monitoring but also increase our knowledge of the mechanisms of AFDAS.

Here, we found the AFDAS group exhibited more cortical involvement compared to the NoAF group, which is conventional considered one of the imaging features in cardioembolic stroke but has not been commonly reported among the AFDAS population in previous studies[21]. Multiple regression analysis revealed a significant correlation between the cortical lesion and AFDAS, indicating that in the absence of significant alternative mechanisms for cerebral infarction, identifying a cortical infarct should prompt more active monitoring for AF.

**Table 3. Multivariate logistic regression model containing lesions distributions clinical variables and final stepwise elimination model between AFDAS and NoAF groups.**

| | OR(95%CI) | p Value |
|---|---|---|
| **Multivariate logistic regression** | | |
| HF | 2.237(0.988-5.572) | 0.053 |
| Hypertension | 0.137(0.025-0.752) | 0.022 |
| Vascular diseases | 0.113(0.024-0.528) | 0.006 |
| Alcohol | 0.670 (0.193-2.321) | 0.527 |
| Prior stroke history | 0.087 (0.010-0.777) | 0.029 |
| CHA2DS2-Vasc score | | |
| <4 | - | 0.539 |
| =4-6 | 1.246(0.255-6.083) | 0.785 |
| >6 | 2.399(0.353-16.317) | 0.371 |
| Cr | 1.040(0.991-1.092) | 0.107 |
| CRP | 1.063(1.021-1.113) | 0.031 |
| Hs-CRP | 1.058(1.013-1.106) | 0.012 |
| BNP | 1.007(1.002-1.013) | 0.006 |
| LAD | 1.462(1.190-1.798) | <0.001 |
| LVD | 1.036(0.873-1.230) | 0.685 |
| Cortical involvement | 1.326(0.215-8.815) | 0.761 |
| Lobe involvement | 2.578(1.217-5.460) | 0.013 |
| Left hemisphere | 8.327(0.531-130.497) | 0.131 |
| Right hemisphere | 13.468(0.779-232.772) | 0.074 |
| Lobe involvement number | | |
| <3 | - | 0.842 |
| =3 | 0.475(0.023-9.881) | 0.631 |
| >3 | 0.869(0.024-30.855) | 0.939 |
| Temporal lobe | 3.358(0.353-31.958) | 0.292 |
| Insular lobe | 1.500(0.131-17.136) | 0.744 |
| **Stepwise elimination model** | | |
| Hypertension | 0.198(0.046-0.853) | 0.030 |
| Vascular diseases | 0.152(0.038-0.606) | 0.008 |
| Prior stroke history | 0.154(0.029-0.823) | 0.029 |
| CRP | 1.051(1.010-1.091) | 0.024 |
| Hs-CRP | 1.047(1.009-1.086) | 0.015 |
| BNP | 1.005(1.001-1.010) | 0.008 |
| LAD | 1.470(1.231-1.757) | <0.001 |
| Temporal lobe | 3.645(1.193-11.137) | 0.023 |

**Table 4.** Multivariate logistic regression model containing lesions distributions clinical variables and final stepwise elimination model between AFDAS and KAF.

|  | OR(95%CI) | p Value |
|---|---|---|
| **Multivariate logistic regression** |  |  |
| Vascular diseases | 0.707(0.232-2.155) | 0.588 |
| Prior stroke history | 0.389(0.108-1.410) | 0.151 |
| Cardiac diseases | 0.507(0.172-1.493) | 0.218 |
| CHA2DS2-Vasc score |  |  |
| < 4 | - | 0.559 |
| =4-6 | 0.714(0.153-3.342) | 0.785 |
| >6 | 0.397(0.061-2.580) | 0.371 |
| GFR | 0.991(0.981-1.044) | 0.517 |
| BNP | 1.000(0.999-1.001) | 0.689 |
| LAD | 0.926(0.850-1.008) | 0.074 |
| LVEF | 0.947(0.965-1.100) | 0.366 |
| Left hemisphere | 3.517(0.072-0.589) | 0.003 |
| Lobes involvement number |  |  |
| <3 | - | 0.018 |
| =3 | 6.994(1.793-27.276) | 0.005 |
| >3 | 2.218(0.346-14.207) | 0.387 |
| Frontal lobe | 0.565(0.165-1.934) | 0.370 |
| **Stepwise elimination model** |  |  |
| LAD | 0.896(0.823-0.976) | 0.014 |
| Left hemisphere | 0.162(0.061-0.425) | <0.001 |
| Number of lobes infarction |  |  |
| =3 | 5.644(1.784-17.861) | 0.003 |

Additionally, more insular and temporal lobes involvements were found in the AFDAS population than in the NoAF group. The lesions located in the insular lobe have frequently been mentioned in previous studies as an important marker of detection of AF after stroke[19-21], as well as considered one of the potential ways to produce post-stroke myocardial damage[22,24,25]. Except been supported by clinical studies, this view has been also determined in animal experiments[18]. The knowledge of temporal lobe involvement in the AFDAS population is very limited, but the patients with temporal lobe atrophy were found to have significantly lower cumulative proportion of 3-year survivors free of dementia[6]. Although in the multiple regression analysis, a significant correlation between temporal lobe involvement and AFDAS was revealed, further and more research is needed to determine whether there is an association between temporal lobe involvement and dementia.

Moreover, the AFDAS population was commonly older than those patients without AF in the previous studies, and hypertension was considered one of the independent risk factors related to AFDAS[13,15,27,28]. Contrary to the previous research, after matching for age, gender, and the time of first stroke onset to hospital admission, the multiple regression analysis performing a stepwise model suggested that a lower proportion of hypertension was significantly associated with AFDAS.

In those patients with a confirmed history of atrial fibrillation before admission, we found most of the lesions located in the left hemisphere, with the majority of patients having bilateral hemispheres involvement. After performing multiple regression analyses with stepwise elimination algorithm, the more left hemisphere involvement was found to be independently associated with KAF. The dominant hemisphere infractions were considered one of independent risk factors for stroke-related dementia and cognitive disorder[29]. Although we did not distinguish the dominant hemisphere in our study, our results still indirectly support this view.

In our study, we found that most KAF patients only had 1 or 2 lobes involved, and the proportion of patients with three lobes involvement was significantly lower than that of the AFDAS group. Additionally, it's interesting to notice that KAF patients had more frontal lobes involvement than AFDAS patients. Lesion located in the frontal lobes has been reported that associated with cognitive disorder in previous studies[29], but unfortunately, we did not evaluate the cognitive function of our patients, so the significance of this difference needs more clinical observation in the future to demonstrate. Moreover, frontal lobe involvement was not independently associated with AFDAS in the multiple regression analysis.

By comparing the AFDAS group with a matched KAF group with respect to age, gender, and the distance between stroke onset and admission, we elucidated the clinical and demographic characteristics and risks in the AFDAS population. Supporting the previous research, we found that the AFDAS group had a lower proportion of peripheral vascular diseases, prior stroke history, cardiac disease history, and lower CHAD2S2-Vasc scores, but they were not the independent risk factors of AFDAS and KAF in the multiple regression analysis.

4.1. Limitations

Our study has several limitations due to the nature of its retrospective observational study design. Additionally, the lack of the admission Holter/ECG results means that the subgroup analysis could not be conducted based on whether the AF was detected on admission. Although long-term cardiac monitoring is typically recommended only for patients who were not diagnosed with AF by admission Holter/ECG, part of patients with asymptomatic, low-burden AF may still be included in the AFDAS patient group. Furthermore, selection bias may be present since the enrollment of patients who underwent long-term cardiac monitoring was not strictly consecutive because patient consent was required before the device was performed. Conventionally, stroke caused by AF is conventionally believed to be more severe than other reasons. However, since the lack of NIHSS scores, let this important indicator for assessing stroke severity was not included in the matching process and intergroup comparison.

5. Conclusions

Here, we found that ischemic stroke patients with atrial fibrillation detected by long-term ECG monitoring have more cortical involvement, lobes involvement, and temporal and insular lobe infraction. Contrary to previous research, after matching for age, gender, and occurrence time from onset to admission, the lower proportion of hypertension, vascular disease, and prior stroke history were independently associated with the AFDAS population. In addition, some biochemical indicators (CRP, Hs-CRP, BNP), LAD, and temporal lobe involvement were identified as independent risk factors for AFDAS by performing a backward stepwise elimination algorithm. We also observed the obvious differences between the AFDAS and KAF populations in terms of risk factors, $CHA_2DS_2$-VAScscores, bio-markers, TTE results, and ischemic stroke lesions distribution. Notably, larger LAD, more involvement of the left hemisphere brain lobes, and less involvement of the three lobes were independently associated with AFDAS and KAF. In conclusion, we believe the distribution of lesions could be a valuable indicator to increase our knowledge about the classification of patients performed with long-term cardiac monitoring.

## 5. Statements and Declarations

*5.1 Competing Interests*

The authors have no relevant financial or non-financial interests to disclose.

*5.2 Ethics declarations*

This is a retrospective, observational study that was conducted on already available data; and we have checked with the ethical committee of Beijing Tiantan Hospital to make sure that ethical approval is not needed in this study and complying with the requirements of China.

*5.3 Author Contributions*

Yiming Chen, MD, performed the data analyses and wrote the manuscript; Sihui Wang, MD, is responsible for the collection and analysis of image data. Dong Xu, MD, Ph.D. contributed to the conception of the study and helped perform the analysis through constructive discussions.